\documentclass[a4paper,prl,preprint,floatfix,superscriptaddress]{revtex4}
\usepackage[english]{babel}
\usepackage{amsmath,amsfonts,amssymb,float}
\usepackage[hidelinks=true]{hyperref}
\usepackage{graphicx,type1cm,eso-pic,color} 
\makeatletter
\newcommand{\watermark}[1]{
\AddToShipoutPicture{%
            \setlength{\@tempdimb}{.5\paperwidth}%
            \setlength{\@tempdimc}{.5\paperheight}%
            \setlength{\unitlength}{1pt}%
            \put(\strip@pt\@tempdimb,\strip@pt\@tempdimc){%
        \makebox(0,0){\rotatebox{45}{\textcolor[gray]{0.85}%
        {\fontsize{3cm}{3cm}\selectfont{#1}}}}%
            }%
}}
\makeatother

\begin{document}

\title{Essential criteria for efficient pulse amplification via Raman and
  Brillouin scattering}

\author{R.M.G.M. Trines}
\affiliation{Central Laser Facility, STFC Rutherford Appleton Laboratory, Didcot, OX11 0QX, United Kingdom}
\author{E.P. Alves}
\affiliation{GoLP/IPFN, Instituto Superior T\'ecnico, Universidade de Lisboa, 1049-001 Lisbon, Portugal}
\affiliation{SLAC National Accelerator Laboratory, Menlo Park, CA 94025, USA}
\author{E. Webb}
\affiliation{Central Laser Facility, STFC Rutherford Appleton Laboratory, Didcot, OX11 0QX, United Kingdom}
\author{J. Vieira}
\affiliation{GoLP/IPFN, Instituto Superior T\'ecnico, Universidade de Lisboa, 1049-001 Lisbon, Portugal}
\author{F. Fi\'uza}
\affiliation{SLAC National Accelerator Laboratory, Menlo Park, CA 94025, USA}
\author{R.A. Fonseca}
\affiliation{GoLP/IPFN, Instituto Superior T\'ecnico, Universidade de Lisboa, 1049-001 Lisbon, Portugal}
\affiliation{DCTI/ISCTE Lisbon University Institute, 1649-026 Lisbon, Portugal}
\author{L.O. Silva}
\affiliation{GoLP/IPFN, Instituto Superior T\'ecnico, Universidade de Lisboa, 1049-001 Lisbon, Portugal}
\author{J. Sadler}
\author{N. Ratan}
\author{L. Ceurvorst}
\author{M.F. Kasim}
\affiliation{Department of Physics, University of Oxford, Oxford OX1 3PU, United Kingdom}
\author{M. Tabak}
\affiliation{Lawrence Livermore National Laboratory, Livermore, CA 94550-9234, USA}
\author{D. Froula}
\author{D. Haberberger}
\affiliation{Laboratory for Laser Energetics, 250 East River Road, Rochester, NY 14623-1299, USA}
\author{P.A. Norreys}
\affiliation{Department of Physics, University of Oxford, Oxford OX1 3PU, United Kingdom}
\affiliation{Central Laser Facility, STFC Rutherford Appleton Laboratory, Didcot, OX11 0QX, United Kingdom}
\author{R.A. Cairns}
\affiliation{University of St Andrews, St Andrews, Fife KY16 9AJ, United Kingdom}
\author{R. Bingham}
\affiliation{Central Laser Facility, STFC Rutherford Appleton Laboratory, Didcot, OX11 0QX, United Kingdom}
\affiliation{SUPA, Department of Physics, University of Strathclyde, Glasgow, G4 0NG, United Kingdom}
\date\today

\begin{abstract}
Raman and Brillouin amplification are two schemes for amplifying and
compressing short laser pulses in plasma. Analytical models have
already been derived for both schemes, but the full consequences of
these models are little known or used. Here, we present new criteria
that govern the evolution of the attractor solution for the seed pulse
in Raman and Brillouin amplification, and show how the initial laser
pulses need to be shaped to control the properties of the final
amplified seed and improve the amplification efficiency.
\end{abstract}

\maketitle

Raman (and later Brillouin) scattering and amplification were first
discovered in solid-state physics \cite{cvraman} and also found
applications in gases and molecular vibrations
\cite{armstrong62,sen65,flora75, kaup79,menyuk92} as well as
non-linear optics \cite{lamb1,lamb2}. Raman and Brillouin scattering
have also been demonstrated in laser-plasma interaction, see
e.g. Forslund \emph{et al.}  \cite{forslund}. Plasma-based
compression and amplification of laser pulses via Raman or Brillouin
scattering has been proposed to overcome the intensity limitations
posed by solid-state optical systems
\cite{maier66,milroy77,milroy79,capjack82,sutyagin,shvets99,
  andreev06}. In the context of Raman or Brillouin amplification,
analytical models have been derived under the assumption that the
basic shape of the growing seed pulse does not change during
amplification, while its amplitude and duration evolve according to
well-defined scaling laws \cite{shvets99,andreev06}. For Raman
amplification (or Brillouin amplification in the weak-coupling
regime), this assumption is only correct if the seed pulse has the
following properties: (i) pulse amplitude is proportional to the
interaction time $t$
\cite{shvets98,dodin00,malkinpop03,pingprl,dino05}, (ii) pulse
duration is proportional to $1/t$, or bandwidth proportional to $t$
\cite{dodin00,malkinpop03,pingprl,dino05,kim03}, (iii) pulse energy is
proportional to $t$, or inversely proportional to its duration
\cite{shvets00,tsidulko02,cheng05,renpop08, yampolski08,malkinprl07},
(iv) the asymptotic ``$\pi$-pulse'' solution is an attractor solution,
i.e. a ``not quite ideal'' seed pulse will reshape itself into an
approximate $\pi$-pulse shape
\cite{shvets99,malkinpop03,kim03,malkinprl07,malkinpop00,yampolsky04,
  trines11b,lehmann1,lehmann2}, (v) in multi-dimensional simulations
where the pulses have a finite transverse width, the seed pulse
acquires a ``horseshoe'' shape \cite{dodin00,trines11b,mardahl02,
  fraiman02,balakin03,balakin05,hur09,trines11a,lehmann3}. Most of the
above also applies to Brillouin amplification in the strong-coupling
regime \cite{andreev06}, although the scalings for the seed pulse
duration and amplitude with interaction time are different. In this
Letter, we perform the first detailed and systematic study of the full
non-linear evolution of the seed pulse for both Raman and Brillouin
amplification, and derive novel criteria for the optimal shape of the
initial seed pulse before, during and after the interaction, which can
be exploited to guide the design of future experiments and maximize
their efficiency. At present, the properties of the ideal attractor
solution for the seed pulse are not used to optimize the design
of Raman or Brillouin amplification experiments.

We define $a_0$ and $a_1$ to be the scaled envelopes
of pump and seed pulse respectively, $a_{0,1} \equiv 8.55\times
10^{-10} g^{1/2} (I_{0,1} \lambda_{0,1}^2 [\mathrm{Wcm}^{-2} \mu
  \mathrm{m}^2])^{1/2}$, where $g=1$ ($g=1/2$) denotes linear
(circular) polarisation. Let $\omega_0$ and $n_{cr}$ denote the pump
laser frequency and critical density, and $n_e$ and $\omega_{pe}$ the
background electron density and corresponding plasma frequency. The
group velocity of the pump pulse is $v_g = c^2 k_0/\omega_0 =
c(1-n_e/n_{cr})^{1/2}$ and the electron thermal velocity is $v_e =
(k_B T_e/m_e)^{1/2}$.

For Raman amplification, the envelope equations for pump, seed and
plasma wave take the following form \cite{shvets99}:
\begin{align}
\label{eq:env1}
(\partial/\partial t \pm v_g \partial/\partial x) a_{0,1} &= \mp i\Gamma_R
a_{1,0} b^{(*)},\\
\label{eq:env3}
(\partial/\partial t + 3v_e^2 (k/\omega_{pe}) \partial/\partial x) b
&= -i\Gamma_R a_0 a_1^*,
\end{align}
where $\Gamma_R a_0$ denotes the Raman backscattering growth rate in
homogeneous plasma and $b \equiv \alpha_R \delta n_e/n_e$ with $\delta
n_e$ the envelope of the electron density fluctuations driven by the
beating of pump and seed pulses, and $\alpha_R$ to be
determined. Comparing these equations to the envelope equations by
Forslund \emph{et al.}  \cite{forslund} yields $\Gamma_R\alpha_R =
\omega_{pe}^2/(4\omega_0)$ and $\Gamma_R/\alpha_R = c^2
k^2/(4g\omega_{pe})$, where $k \approx 2k_0 \approx 2\omega_0/c$ is
the wave number of the RBS Langmuir wave (for $\omega_{pe} \ll
\omega_0$). Then $\Gamma_R = [\omega_0 \omega_{pe}/(4g)]^{1/2}$ and
$\alpha_R = \sqrt{g}(\omega_{pe}/\omega_0)^{3/2}/2$, with
(\ref{eq:env1})-(\ref{eq:env3}) valid for
$a_0 < a_{wb} = \alpha_R/\sqrt{2}$ \cite{shvets99}.

Following Malkin, Shvets and Fisch \cite{shvets99}, or Menyuk, Levi
and Winternitz \cite{menyuk92}, we
define $\zeta = x/c + t$, $t' = \Gamma_R^2 a_{00}^2 t$
and $\xi = 2\sqrt{\zeta t'}$ where $a_{00}$ denotes the pump pulse
amplitude. Attractor solutions to the above system can then be
obtained in terms of $\xi$ alone. In particular, the first peak of the
growing seed pulse can be approximated by $a_1(\zeta,t') \approx
(2t'/\Gamma_R\xi) \partial u(\xi)/\partial\xi$ where $u(\xi) =
2\sqrt{2}\arctan[\epsilon \exp(\xi)/ (4\sqrt{2\pi\xi})]$, with
$\epsilon < 0.1$ depending on the initial seed pulse B-integral. The
function $\partial u(\xi)/\partial\xi$ has an amplitude $A \approx
1.29$ and a width $\Delta \xi \approx 2.65$, mostly independent of
$\epsilon$, while the position of its maximum, $\xi_M$, obeys $5 < \xi_M
< 7$ for practical values of $\epsilon$ \cite{shvets99,dodin00}. Let
$\Delta\zeta$ denote the width of the first peak of $a_1(\zeta,t')$
for fixed $t'$ and let $\xi_{1,2} = \xi_M \pm \Delta \xi/2$. Then
$\xi_M\Delta \xi = (\xi_2^2 -\xi_1^2)/2 = 2 t' \Delta\zeta$.
For $\xi=\xi_M$, we find that $||a_1|| = 2A t'/(\Gamma_R
\xi_M)$, $\Delta\zeta = \xi_M \Delta \xi/(2 t')$. We consider pump
and seed pulses with durations $\tau_0$ and $\tau_1$ (after
amplification), and setting $\Delta\zeta = \tau_1$ and $t
= \tau_0/2$ (for an interaction time $t$, the counter-propagating seed
pulse consumes $2t$ of pump pulse), we find:
\begin{align}
\label{eq:selfsim1}
\Gamma_R^2 a_{00}^2 \tau_0 \tau_1 &= \xi_M\Delta\xi \approx 15,\\
\label{eq:selfsim2}
\Gamma_R ||a_1|| \tau_1 &= A\Delta\xi \approx 3.4.
\end{align}
The asymptotic energy transfer efficiency for the first peak is then
given by $\eta = ||a_1||^2\tau_1(t)/(2 a_0^2 t) = A^2 \Delta\xi/\xi_M
\approx 4.4/\xi_M$. Thus, $\eta$ is constant for a given
configuration, and decreases with increasing $\xi_M$. We confirm these
predictions in our simulations below.

The purpose of these equations is as follows. Eq. (\ref{eq:selfsim1})
allows one to derive scalings for the seed pulse duration $\tau_1(t)$
and amplitude $a_1(t)$, and also to tune these parameters via the
intensity of the pump pulse \cite{trines11b}. Eq. (\ref{eq:selfsim2})
provides a relationship between seed pulse duration and amplitude that
does not depend on the pump pulse at all (the only combination of
$a_1$ and $\tau_1$ with this property). This is important for the
tailoring of the initial seed pulse in experiments: $\tau_1(0)$ and
$a_1(0)$ are not independent parameters, but should obey Eq.
(\ref{eq:selfsim2}) for optimal energy transfer, otherwise the seed
pulse will first reshape itself and only be amplified after that
\cite{yampolsky04,lehmann1,lehmann2,trines11b}, reducing the
amplification efficiency.

Brillouin scattering in the so-called weak-coupling regime
\cite{milroy79,sutyagin,forslund,cohen79,cohen01,williams} is very
similar to Raman scattering and can be treated in the same way. We
introduce $\omega_{pi} = \omega_{pe}\sqrt{Zm_e/m_i}$ and $c_s =
v_e\sqrt{Zm_e/m_i}$. For $a_{00}^2 < 8g (\omega_0/\omega_{pe})^2 c_s
v_e^2/c^3$, the electron pressure is the dominant restoring force and
the plasma wave dispersion is not significantly affected by the
beating between pump and seed pulses. In that case one can reuse
equations (\ref{eq:env1})-(\ref{eq:env3}) and only needs to replace
$3v_e^2 (k/\omega_{pe})$ by $c_s$ in (\ref{eq:env3}). For backward
Brillouin scattering, the ion-acoustic wave has wave number $k_s=2k_0$
and frequency $\omega_s = c_s k_s = 2c_s k_0$. Then we find
$\Gamma_B\alpha_B = \omega_{pe}^2/ (4\omega_0)$ and $\Gamma_B/\alpha_B
= c^2 c_s^2 k_s^2/(4g\omega_s v_e^2)$, leading to $\Gamma_B = c
\omega_{pe} \omega_s/(4 v_e \sqrt{g\omega_0 \omega_s})$ and $\alpha_B
= \sqrt{g} v_e \omega_{pe} /(c\sqrt{\omega_0 \omega_s})$.  After
substituting $\Gamma_B$, $\alpha_B$ for $\Gamma_R$, $\alpha_R$, all
the above results for Raman amplification also apply to the
weak-coupling Brillouin case, including Eqns. (\ref{eq:selfsim1}) and
(\ref{eq:selfsim2}), the wave breaking threshold $a_{wb} =
\alpha_B/\sqrt{2}$, the numerical constants $5 < \xi_M < 7$,
$\Delta\xi \approx 2.65$ and $A \approx 1.29$, and the seed pulse
scalings.

For $a_{00}^2 > 8g (\omega_0/\omega_{pe})^2 c_s v_e^2/c^3$ or
$\Gamma a_0/\omega_0 > c_s/c$, the ponderomotive pressure
from the beating between pump and seed pulses will take over from the
thermal pressure as the primary restoring force for the ion-acoustic
wave. In this regime, called \emph{strong-coupling} (sc) Brillouin
scattering, the equation for the plasma wave becomes \cite{andreev06}:
\begin{equation}
\label{eq:scplasma}
\partial^2 b /\partial t^2  = - \alpha_{sc} c^2 k^2
\omega_{pi}^2 / (2g\omega_{pe}^2) a_0 a_1 = -\Gamma_{sc}^2 a_0 a_1.
\end{equation}
From (\ref{eq:env1}) and (\ref{eq:scplasma}) and using $k = 2k_0 =
2\omega_0 v_g/c^2$ as before, we find: $\Gamma_{sc}\alpha_{sc} =
\omega_{pe}^2/(4\omega_0)$ and $\ \Gamma_{sc}^2/\alpha_{sc} = c^2 k^2
\omega_{pi}^2 / (2g\omega_{pe}^2)$.  This yields
\cite{forslund,huller,andreev06}:
\begin{align}
\Gamma_{sc} &= [(v_g/c)^2 \omega_{pi}^2 \omega_0/(2g)]^{1/3} =
(2\omega_s \Gamma_B^2)^{1/3}, \\
\Omega_{sc} &= \omega_{sc} +i\gamma_{sc} = [(1+i\sqrt{3})/2]\Gamma_{sc} a_{00}^{2/3},
\end{align}
where $\omega_{sc}$ and $\gamma_{sc}$ denote the frequency and growth
rate of the ion-acoustic wave. Following the approach by Andreev
\emph{et al.}  \cite{andreev06}, we define $t' = \Gamma_{sc}
a_{00}^{2/3} t$, $\zeta = \Gamma_{sc}a_{00}^{2/3} (t+x/v_g)$ and $\xi
= \zeta\sqrt{t'}$. Again, attractor solutions to the system
(\ref{eq:env1}) and (\ref{eq:scplasma}) can be obtained in terms of
$\xi$ alone. In particular, the resulting seed pulse will scale as
$a_1(\zeta,t') = a_{00} (t')^{3/4} f(\xi)$ where $f(\xi)$
has a fixed duration $\Delta\xi \approx 3.3$ and an amplitude $A \approx
0.62$ \cite{andreev06,lehmann2}. Using $\Delta\xi =
(\Delta\zeta)\sqrt{t'}$ for fixed $t'$, $||a_1|| = a_{00} A
(\Delta\xi/\Delta\zeta)^{3/2}$ and inserting $\Delta\zeta
=\Gamma_{sc} a_{00}^{2/3}\tau_1$ and $t' = \Gamma_{sc} a_{00}^{2/3}
\tau_0/2$ for pump and seed pulses with durations $\tau_0$ and
$\tau_1$ into $\Delta\xi$ yields:
\begin{align}
\label{eq:ssbril1}
\Gamma_{sc}^3 a_{00}^2 \tau_0 \tau_1^2 &= 2(\Delta\xi)^2 \approx 22,\\
\label{eq:ssbril2}
\Gamma_{sc}^3 ||a_1||^2 \tau_1^3 &= A^2 (\Delta\xi)^3 \approx 13.8.
\end{align}
The scaling for the seed pulse duration is then $\tau_1(t) =
\Delta\xi/(\Gamma_{sc}^{3/2}a_{00} t^{1/2})$, and the asymptotic
efficiency is $\eta = a_1^2(t)\tau_1(t)/(2a_{00}^2 t) = A^2 \Delta\xi/2
\approx 0.63$. The role of (\ref{eq:ssbril1}) and (\ref{eq:ssbril2})
matches that of (\ref{eq:selfsim1}) and
(\ref{eq:selfsim2}) for Raman amplification.

To verify the validity of Eqns. (\ref{eq:selfsim2}) and
(\ref{eq:ssbril2}), we have carried out one-dimensional
particle-in-cell simulations using the codes XOOPIC \cite{xoopic} and
OSIRIS \cite{osiris}. We used a long pump laser beam with constant
amplitude $a_0$ and wave length $\lambda_0 = 1\ \mu$m ($\omega_0 =
2\pi c/\lambda_0$, $n_{cr} = \varepsilon_0 m_e \omega_0^2/e^2$), a
long plasma column with constant electron density $n_0$ and plasma
frequency $\omega_{pe}$, and a seed pulse with initial amplitude
$a_1(0) = a_0$ and duration $\tau_1(0)$. For the Raman simulations, we
used a plasma density corresponding to $\omega_{pe}/\omega_0 = 1/15$,
pump laser amplitudes $a_{wb}/4$, $a_{wb}/2$, $3a_{wb}/4$, $a_{wb}$
and $2a_{wb}$, with $a_{wb}$ adjusted for each plasma density, and
pump pulse durations up to $2\times 10^5 /\omega_0 \approx 112$
picoseconds. We use $\tau_1(0)/\tau_R = 0.1$, 0.5, 1.0 and 2.0, where
$\tau_R [\mathrm{s}] = 4.22\times 10^{-6} \lambda_0 [\mu\mathrm{m}]
(n_e/n_{cr})^{-1/4} (I_1 \lambda_1^2 [\mathrm{Wcm}^{-2} \mu
  \mathrm{m}^2])^{-1/2}$ is taken from (\ref{eq:selfsim2}). For the
Brillouin simulations, we used $m_i/(Zm_e) = 1836$, a plasma density
$n_e = 0.3 n_{cr}$ and pump amplitudes $a_0= 0.0085$, 0.027 and 0.085,
corresponding to $10^{14}$, $10^{15}$ and $10^{16}$ W cm$^{-2}$, and
pump pulse durations of 11.4 ps, 3.8 ps and 1.1 ps respectively. We
use $\tau_1(0)/\tau_B = 0.1$, 0.2, 0.5, 1.0, 2.0 and 5.0, where
$\tau_B [\mathrm{s}] = 1.78\times 10^{-9} \lambda_0 [\mu\mathrm{m}]
[(Zm_e/m_i) (n_e/n_{cr})(1-n_e/n_{cr}) (I_1 \lambda_1^2
  [\mathrm{Wcm}^{-2} \mu\mathrm{m}^2])]^{-1/3} $ is taken from
(\ref{eq:ssbril2}). The parameters of the simulations are discussed at
length in the Supplementary Information \cite{supp}.

\begin{figure}[ht]
\includegraphics[width=0.45\textwidth]{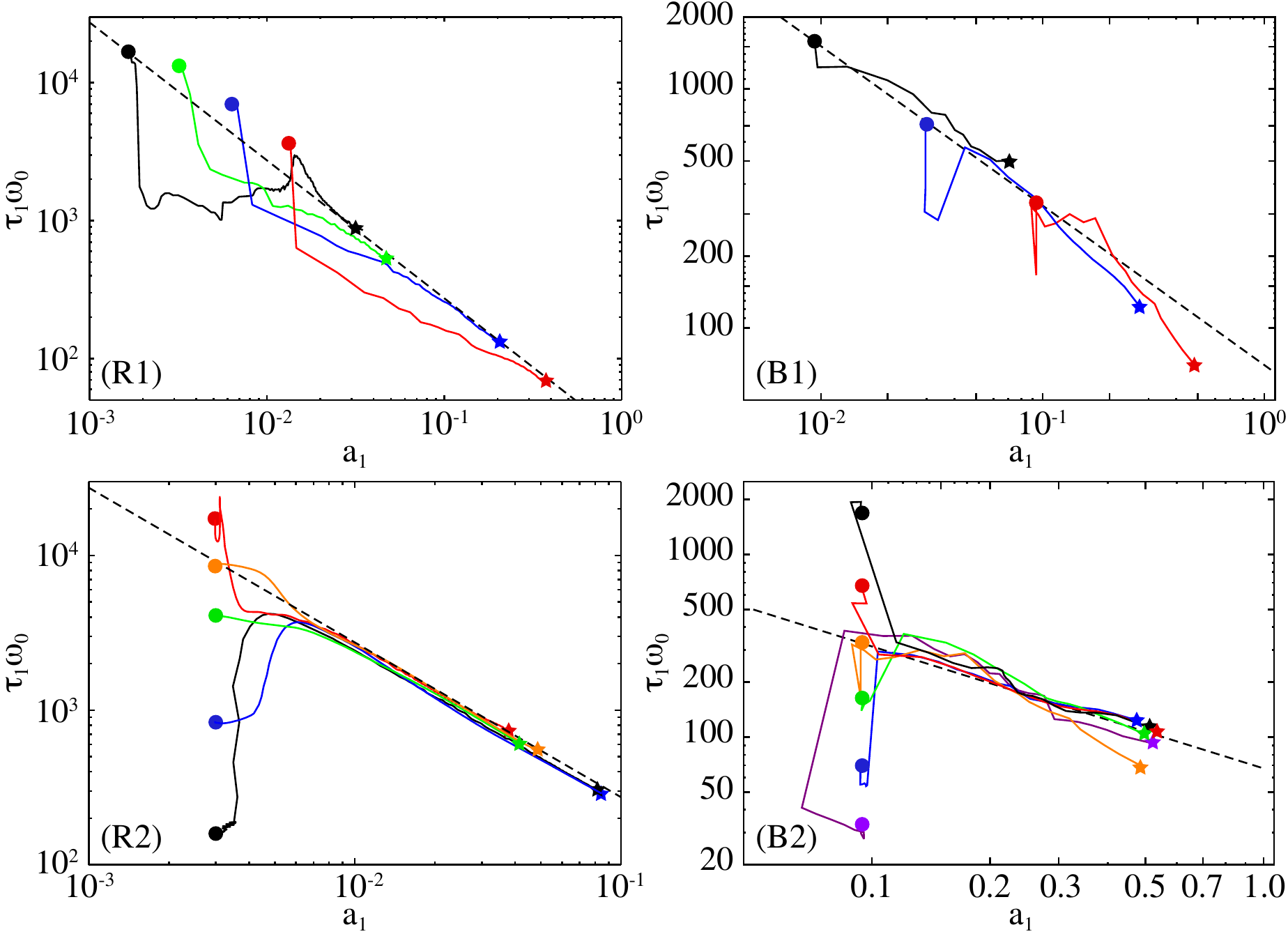}
\caption{Evolution of duration ($\tau_1$) versus peak amplitude
  ($a_1$) of Raman (R1,R2) and sc-Brillouin (B1,B2) amplified pulses
  for different pump amplitudes (R1,B1) or initial seed durations
  (R2,B2), demonstrating the attractor nature of the ideal
  solutions. R1: $n_0/n_{cr} = 0.0044$, and $a_0/a_{wb}=0.25$ (black),
  0.5 (green), 1.0 (blue) and 2.0 (red), with initial seed amplitude
  $a_1 = a_0$ for each case. B1: $n_0/n_{cr} = 0.3$, and $a_0=0.0085$
  (black), 0.027 (blue) and 0.085 (red); again, $a_1 = a_0$. R2:
  $n_0/n_{cr} = 0.0044$, $a_1 = a_0 =0.75 a_{wb}$. The initial
  durations of the seed pulse are $\tau_1/\tau_R = 0.02$ (black), 0.1
  (blue) 0.5 (green), 1.0 (orange) and 2.0 (red). B2: $n_0/n_{cr} =
  0.3$, and pump and seed amplitudes of $a_0 = a_1 = 0.085$. The
  initial durations of the seed pulse are $\tau_{1}/\tau_{B}=0.1$
  (purple), 0.2 (blue), 0.5 (green), 1.0 (orange), 2.0 (red), and 5.0
  (black). The dashed lines correspond to Eqns. (\ref{eq:selfsim2})
  for Raman and (\ref{eq:ssbril2}) for sc-Brillouin respectively.}
\label{fig:1}
\end{figure}

In Figure \ref{fig:1}, we show the evolution of $\omega_0\tau_1$
versus $a_1$ for simulations of Raman (left) and Brillouin (right)
amplification, to demonstrate the ``attractor'' nature of the optimal
seed pulse solutions (\ref{eq:selfsim2}) and (\ref{eq:ssbril2}). The
dashed lines in each frame represent Eqns. (\ref{eq:selfsim2}) and
(\ref{eq:ssbril2}) evaluated for these cases. The top two frames show
($\omega_0\tau_1$,$a_1$) for various initial pump and seed pulse
intensities, where $\tau_1(0) = \tau_R$ or $\tau_1(0) = \tau_B$ in
each simulation. We find that the evolving seed pulses closely follow
the analytical predictions, irrespective of the pump intensity chosen
in the simulations, proving that the predictions by
(\ref{eq:selfsim2}) and (\ref{eq:ssbril2}) for Raman or Brillouin
amplification represent ``attractor'' solutions and remain valid over
a wide range of pulse intensities. For $a_1(t)$ and $\tau_1(t)$, we
find that $a_1(t)\propto t$ and $\tau_1(t) \propto 1/t$ for Raman
amplification and $a_1(t)\propto t^{3/4}$ and $\tau_1(t) \propto
t^{-1/2}$ for sc-Brillouin, although minor aberrations from these
scalings were found at the highest pulse intensities due to non-linear
effects not covered by the three-wave models, e.g. when the seed pulse
becomes powerful enough to drive a wakefield. The bottom two frames
show ($\omega_0\tau_1$,$a_1$) for fixed pulse intensities, while the
initial pulse duration was moved away from the analytical predictions.
We find that in each case the seed pulse first evolves until the pair
($\omega_0\tau_1$,$a_1$) matches (\ref{eq:selfsim2}) or
(\ref{eq:ssbril2}), and then amplifies as dictated by these equations.
This specific behaviour was found in all our simulation results,
irrespective of the plasma density or pump pulse intensity we
used. This proves the following: (i) the $\pi$-pulse solution for
Raman and its Brillouin equivalent are indeed attractors, as predicted
\cite{shvets99,andreev06}, and (ii) changing the initial seed pulse
duration has no significant effect on the end result, so $\tau_1(0)$
should not be treated as a free parameter. The free parameters for
both Raman and Brillouin amplification are the pump wave length
$\lambda_0$ and the density ratio $n_0/n_{cr}$; once these two are
chosen, the position of the attractor $(\tau_1,a_1)$ curve is
completely determined by (\ref{eq:selfsim2}) or
(\ref{eq:ssbril2}). The intensities of the pulses determine the speed
at which the seed pulse evolves, but not the trajectory of
$(\tau_1(t),a_1(t))$.

\begin{figure}[ht]
\includegraphics[width=0.45\textwidth]{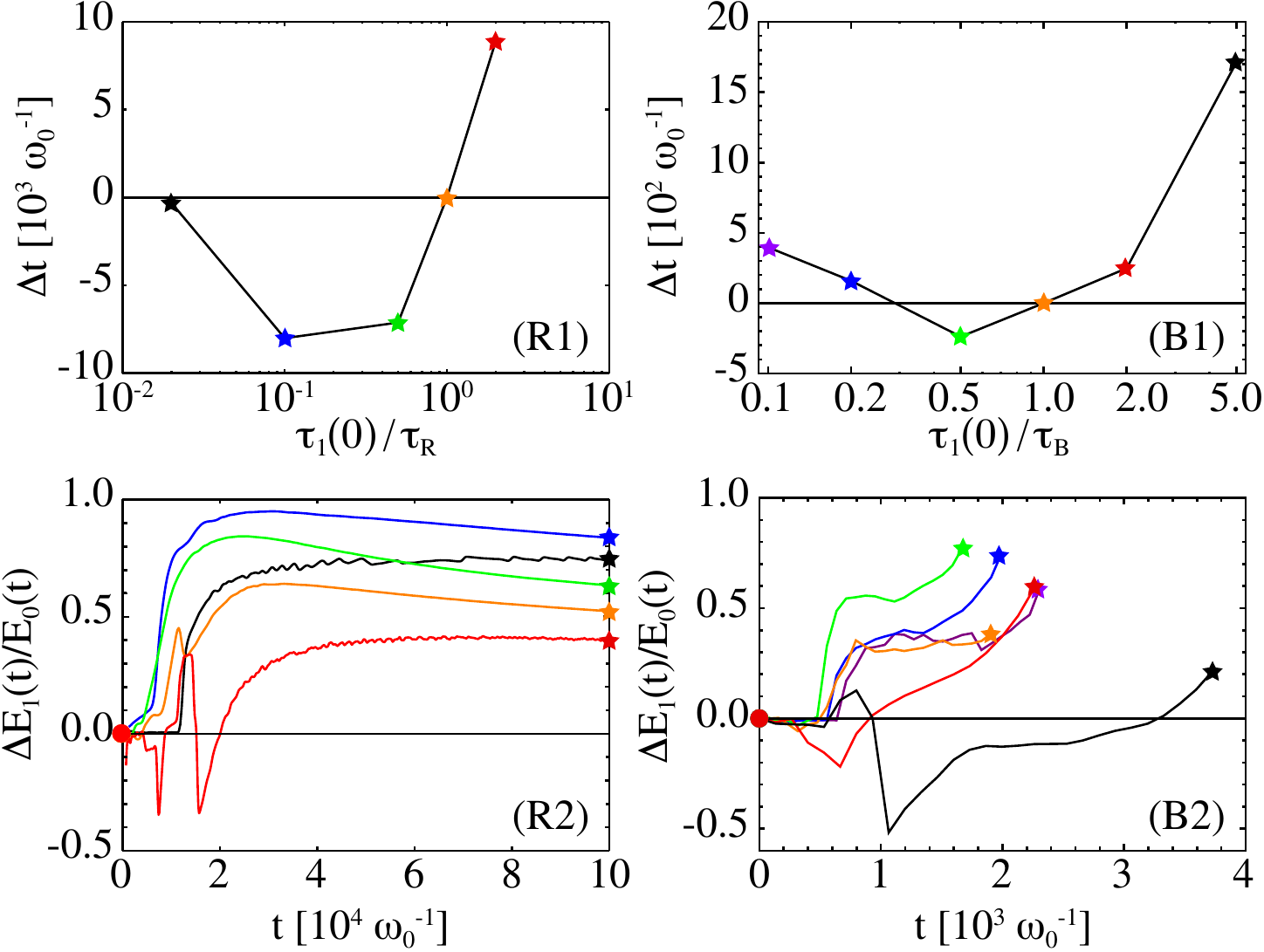}
\caption{(R1,B1): temporal delay $\Delta t$ in reaching a given
  intensity amplification level $(25\times I_\mathrm{pump})$ for seed pulses
  with various initial durations, for the same cases as shown in
  frames (R2,B2) of Figure \ref{fig:1}. (R2,B2): efficiency of the
  amplification process for these same cases; $\Delta E_1$ is the seed
  energy gain, $E_0$ is the absorbed pump energy.}
\label{fig:2}
\end{figure}

In Figure \ref{fig:2}(R1,B1), we show the time needed for seed pulses
with different initial durations to reach a given intensity
$(25\times I_\mathrm{pump})$, for the same cases as shown in frames
(R2,B2) of Figure \ref{fig:1}. We find that amplification is optimal
when $0.2 < \tau_1(0)/\tau_{R,B} < 0.5$, ($\tau_{R,B}$ given by
(\ref{eq:selfsim2}) or (\ref{eq:ssbril2})) while significant delays are
incurred for $\tau_1(0)/\tau_{R,B} > 1$ or $< 0.2$. In Figure
\ref{fig:2}(R2,B2) we show the efficiency of the amplification process
for these same cases ($\Delta E_1 \equiv E_1(t) - E_1(0)$, so $\Delta
E_1 < 0$ means that the seed pulse is losing energy rather than
gaining). Unsurprisingly, a longer delay in amplification is always
accompanied by a lower efficiency. For Raman amplification in
particular, we also find that (i) the asymptotic efficiency is mostly
constant, (ii) the cases showing the longest delay also exhibit the
lowest asymptotic efficiency. This corresponds to the notion that the
Raman efficiency is $\eta \approx 4.4/\xi_M$ (see above) and that
$\xi_M$ increases for non-ideal initial seed pulses that incur longer
delays, in line with predictions for $\xi_M$ in
Ref. \cite{shvets99}. So a poorly chosen initial seed pulse duration
will affect the entire amplification process, not just the initial
stages. Also, the longest delays correspond to an interaction
length of several mm, longer than what is used in many
experiments \cite{ren07,renpop08,kirkwood07,ping09}. This highlights
the need to choose the initial seed pulse parameters according to
Eqns. (\ref{eq:selfsim2}) and (\ref{eq:ssbril2}).

\begin{figure}[ht]
\includegraphics[width=0.45\textwidth]{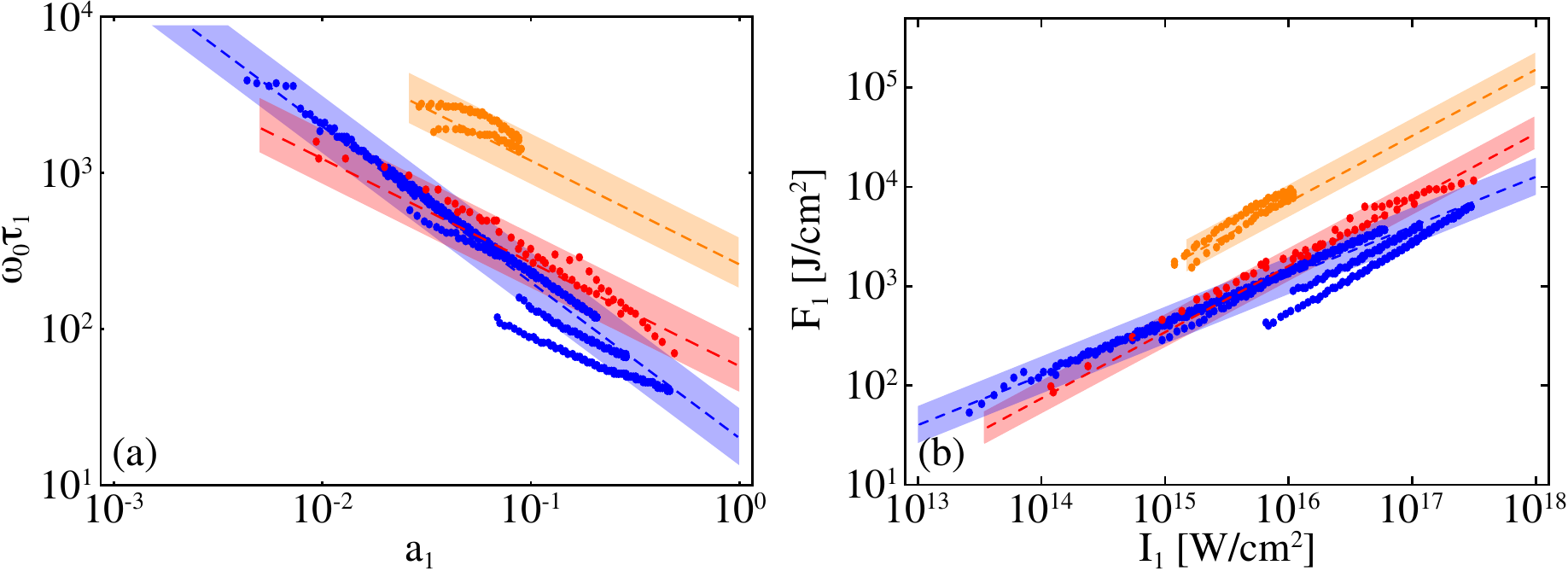}
\caption{Left: map of $(a_1,\tau_1)$ of successfully amplified
  seed pulses via Raman amplification (blue), sc-Brillouin
  amplification at over-quarter-critical densities (red) and
  sc-Brillouin amplification at sub-quarter-critical densities
  (orange). The points are taken from 1-D PIC simulations while the
  shaded areas indicate predictions by Eqns. (\ref{eq:selfsim2}) and
  (\ref{eq:ssbril2}). Right:
  The same data shown on the left is presented in a map of seed
  fluence versus seed intensity, for 1 $\mu$m pump pulse wave
  length.
}
\label{fig:3}
\end{figure}

To make the connection with past Raman or Brillouin amplification
experiments, we have applied our findings to the initial conditions of
various Raman \cite{ren07,renpop08,kirkwood07, ping09} and Brillouin
amplification experiments \cite{lancia10,lancia16}. For the initial
seed pulses in the Raman experiments, we find that $\Gamma_R a_1
\tau_1 = 0.13$ \cite{ren07,renpop08}, 0.22 \cite{kirkwood07}, or 0.86
\cite{ping09}, well below our value of 3.4. For the Brillouin
experiments, we find that $\Gamma_{sc}^3 a_1^2 \tau_1^3 = 8.6$
\cite{lancia10} or 1.31 \cite{lancia16} again below our value of
13.8. Interestingly, the output pulse after the first pass of the
experiments by Ren \emph{et al.}  \cite{ren07,renpop08} has $\Gamma_R
a_1 \tau_1 \approx 3.5$, matching our predictions. Thus, future
experiments need either more powerful seed pulses (up to two orders of
magnitude extra power for Raman and one order for Brillouin, for
similar duration) or longer interaction distances, to allow the pulses
time to reshape themselves. We also note that the experiments with the
biggest energy gain, Refs. \cite{ren07,lancia10} are also the ones
using pulses that match our predictions most closely.

In Figure \ref{fig:3}, we compare 1-D simulations of Raman
amplification at $0.0025 < n_0/n_{cr} < 0.01$ (blue) and sc-Brillouin
amplification at $m_i/(Zm_e) = 1836$, $0.275 < n_0/n_{cr} = 0.325$
(red) and $0.0075 < n_0/n_{cr} = 0.0125$ (orange). The density ranges
have been chosen to minimize the impact of unwanted instabilities
\cite{trines11a}. The dots mark the simulation results, while the
shaded areas mark the predictions by Eqns. (\ref{eq:selfsim2}) and
(\ref{eq:ssbril2}). The left frame shows $\tau_1$ versus $a_1$ for the
amplified seed; the right frame shows the energy flux $F_1 =
I_1\tau_1$ versus intensity $I_1$ for the same cases. All simulation
points lie within the theoretically predicted shaded regions,
highlighting the robustness of (\ref{eq:selfsim2}) and
(\ref{eq:ssbril2}) for a broad range of parameters. Note that
Raman and high-density sc-Brillouin produce the shortest pulses
and highest intensities.  Low-density Brillouin amplification reaches
lower peak intensities but yields the highest pulse fluence because of
longer pulse durations. These results serve as an important guide when
choosing not only the laser and plasma parameters, but also the
preferred amplification scheme when designing an experiment to obtain
a desired output pulse.

Finally, we applied our findings to the results of previously
published 2-D Raman and Brillouin simulations
\cite{trines11a,vieira16}, where both the seed pulse amplitude and
duration depend on the transverse coordinate $r$. We found that even
if $a_1(r)$ and $\tau_1(r)$ depend on $r$ individually, they still
obey Eqns. (\ref{eq:selfsim2}) and (\ref{eq:ssbril2}). For example,
for seed pulses with a Gaussian envelope,
$a_1(x_2)\propto\exp(-r^2/w_0^2)$, this results in a horseshoe
envelope $\tau_1(r) \propto \exp(r^2/w_0^2)$, as seen in Refs.
\cite{dodin00,trines11b,mardahl02,fraiman02,balakin03,
  balakin05,hur09,trines11a,lehmann3}. For donut-shaped seed pulses
with orbital angular momentum, this even results in a ``double
horseshoe'' envelope \cite{vieira16}. These findings are discussed at
length in the Supplementary Material \cite{supp}

We have explored the full non-linear evolution of the seed pulse in
Raman and Brillouin amplification, and derived essential criteria
governing $a_1(t)$ and $\tau_1(t)$, in particular specific products of
$a_1$ and $\tau_1$ that are independent of the pump pulse
properties. We have proved the validity of these criteria in 1-D and
2-D particle-in-cell simulations. Furthermore, we have demonstrated
the importance of choosing the initial seed duration wisely: a
non-optimal value for this parameter (far from the ideal attractor
solution) will delay amplification of the seed pulse and reduce
efficiency. In relation to experiments on parametric amplification,
our results provide unique criteria for their design, and novel
predictions for the properties of the amplified seed pulse, and advice
on which scheme to choose to obtain the desired end result. Since the
ideal amplified seed pulse assumes the shape of a cnoidal wave
\cite{armstrong62,shvets99,andreev06}, our results also explain the
``bursty behaviour'' observed in Brillouin scattering
\cite{weber06,weber11}, as the scattered radiation assumes a very
similar shape. Since the equations for Raman scattering in solid-state
physics or non-linear optics have a shape similar to
Eqns. (\ref{eq:env1})-(\ref{eq:env3}) (see
Refs. \cite{armstrong62,sen65,flora75,kaup79,menyuk92,lamb1,lamb2})
our results will be useful for Raman and Brillouin scattering in
general (not just in plasma), or to any optical three-wave process
with Kerr or $\chi^{(3)}$ non-linearity and counter-propagating
pulses, ensuring a wide range of applications.

\begin{acknowledgments}
This work has been carried out within the
framework of the EUROfusion Consortium and has received funding from
the Euratom research and training programme 2014-2018 under grant
agreement No. 633053. The views and opinions expressed herein do not
necessarily reflect those of the European Commission. The authors
acknowledge financial support from STFC, from the European Research
Council (ERC-2010-AdG Grant 167841), from FCT (Portugal) grant
No. SFRH/BD/75558/2010 and from LaserLab Europe, grant no. GA
654148. We acknowledge PRACE for providing access to resources on
SuperMUC (Leibniz Supercomputing Centre, Garching, Germany).
\end{acknowledgments}

\ \\
\emph{Parameters of the numerical simulations.}
For the simulations in the main manuscript, we have used the
particle-in-cell codes XOOPIC \cite{xoopic} and OSIRIS
\cite{osiris}. The parameters are discussed in detail here. We
distinguish numerical parameters (spatial resolution, time step,
number of particles per grid cell) and physical parameters (laser
pulse duration, spot diameter and amplitude, plasma density, plasma
species, laser-plasma interaction length, etc.)

Both the Raman and Brillouin runs that have been performed with for
figures 1 and 2 of maion manuscript have been done using a moving
simulation window. This window followed the seed pulse, while the pump
pulse was brought into the simulation box via a time-dependent
boundary condition.

The numerical parameters were as follows. For the Raman runs (frames
R1 and R2 in both figures of the main manuscript), the spatial
resolution was 50 points per pump laser wavelength (i.e. $dx = 21$
nm). The time step was given by $dt = 0.95dx/c$. The number of
particles was 100 particles per cell per species. The interpolation
between particles and grid was done using quadratic splines. Ions were
treated as an immobile background. For the Brillouin runs (frames
B1 and B2 in both figures of the main manuscript), the spatial
resolution was $dx = 0.5\lambda_D$, where $\lambda_D$ is the Debye
length. This corresponds to about 220 points per pump laser
wavelength (i.e. $dx = 4.8$ nm). The time step was again $dt =
0.95dx/c$. The number of particles was again 100 particles per cell
per species, and cubic splines were used for interpolation.

The physical parameters were as follows. The pump laser wave length
was 1 $\mu$m for both the Raman and Brillouin simulations. The seed
laser wave length for the Raman simulations was 1.07 $\mu$m, chosen to
ensure that $\omega_1 = \omega_0 - \omega_{pe}$. The seed laser wave
length for the Brillouin runs was 1 $\mu$m. (This hardly matters since
the frequency difference between pump and seed pulses in Brillouin
amplification is considerably less than the seed pulse bandwidth.) For
the Brillouin runs in figure 1, frames B1, the interaction length was
about $2\times 10^4 c/\omega_0$, $8\times 10^3 c/\omega_0$, and
$2\times 10^3 c/\omega_0$ for the simulations with pump intensity
$10^{14}$, $10^{15}$ and $10^{16}$ W/cm$^2$ simulations of frame
B1. The interaction lengths for the simulations in figure 1, frame B2,
are $2\times 10^3 c/\omega_0$ in each case. This corresponds to an
interaction distance of 335 micron, or a 2.2 ps pump pulse duration.
The interaction distance for the Raman runs in figure 1 was up to
$10^5 c/\omega_0$ or up to $16$ mm in each case.

The interaction distance for the simulations displayed in figure 2 is
up to $10^5 c/\omega_0$ for the Raman runs, and up to $2\times 10^3
c/\omega_0$ for the Brillouin runs.

The plasma density was chosen to be $n_e = n_{cr}/225$ for the Raman
simulations, and $n_e = 0.3n_{cr}$ for the Brillouin simulations. The
plasma electron temperature was chosen to be 1 eV for the Raman
simulations and 500 eV for the Brillouin simulations.

\ \\
\emph{Transverse effects.}
In a multi-dimensional setting, the amplitude $a1$ and duration
$\tau_1$ of the growing seed laser pulse will of course depend on the
transverse coordinate $x_2$. The same holds true for the location of the
seed pulse maximum, $\tau_M$. This leaves an imprint on the full shape
of the envelope when parametric amplification is studied in more than
one dimension.

From Equations (\ref{eq:selfsim2}) and (\ref{eq:ssbril2}), 
we find that the $x_2$-dependence of the amplitude $a_1$ also induces an
$x_2$-dependence of the duration $\tau_1$, even for fixed $t$. For example,
for seed pulses with a Gaussian envelope, $a_1(x_2)\propto\exp(-x_2^2/w_0^2)$,
this results in a horseshoe envelope $\tau_1(x_2) \propto \exp(x_2^2/w_0^2)$
\cite{dodin00,trines11b,mardahl02,fraiman02,balakin03,
  balakin05,hur09,trines11a,lehmann3}. For donut-shaped seed pulses
with orbital angular momentum, this even results in a ``double horseshoe''
envelope \cite{vieira16}. We will verify such horseshoe seed pulse shapes
against Eqns. (\ref{eq:selfsim2}) and (\ref{eq:ssbril2}) here.

\begin{figure}[ht]
\includegraphics[width=0.45\textwidth]{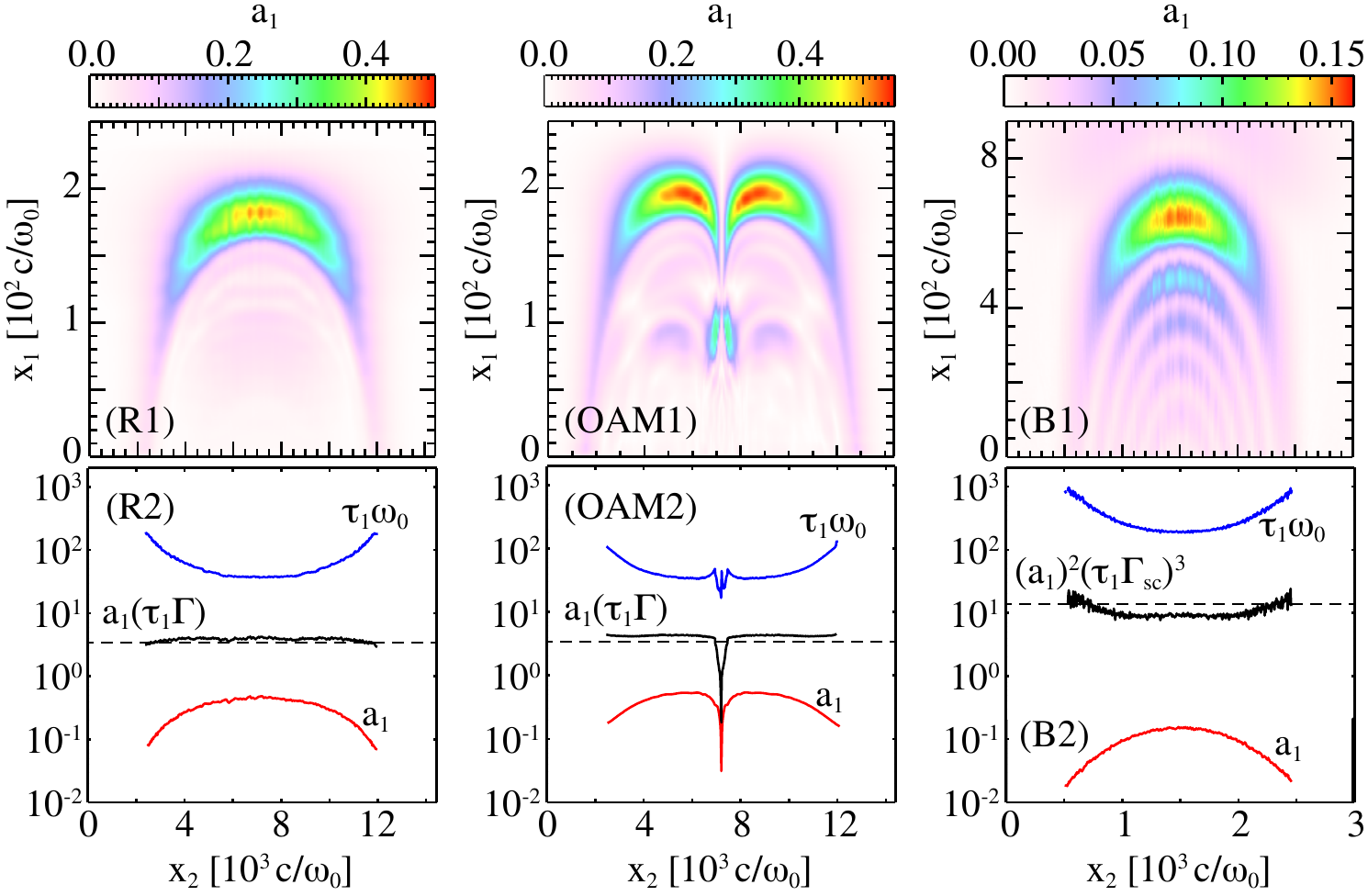}
\caption{Analyzing the curved shape of the amplified seed pulse for
  Raman (R) and Raman with OAM (OAM) and Brillouin (B) amplification
  in multi-dimensional simulations. The top frames show the amplitude
  envelopes of the amplified seeds. The bottom frames show $a_1(x_2)$
  (red) and $\omega_0\tau_1(x_2)$ (blue) versus the transverse
  coordinate $x_2$. The products $a_1(x_2)\Gamma_R\tau_1(x_2)$ (R,OAM)
  and $a_1^2(x_2)[\Gamma_{sc}\tau_1(x_2)]^3$ (B), given by black
  curves, are then verified against Eqns.(\ref{eq:selfsim2}) and
  (\ref{eq:ssbril2}) (dashed lines).}
\label{fig:4}
\end{figure}

In Figure \ref{fig:4}, we verify the horseshoe shape of seed pulses
after amplification. Here, we present the analysis of a 2-D Raman
simulation (R), taken from Ref. \cite{trines11a}, Figure 2a, a 3-D
Raman simulation with orbital angular momentum and a doughnut-shaped
seed pulse intensity envelope (OAM), taken from Ref. \cite{vieira16},
Figure 1a and 2a, and a 2-D sc-Brillouin simulation (B), using
$m_i/(Zm_e) = 1836$, $n_0/n_{cr} = 0.3$ and $a_0=0.027$. Frames R1,
OAM1 and B1 show the vector potential envelopes for these
pulses. Frames R2, OAM2 and B2 show $a_1(x_2)$, $\omega_0\tau_1(x_2)$
and $\Gamma_R a_1(x_2)\tau_1(x_2)$ and $A\Delta\xi \approx 3.4$
(R2,OAM2) or $\Gamma_{sc}^3 a_1^2(x_2) \tau_1^3(x_2)$ and $A^2
\Delta\xi^3 \approx 13.8$ (B2), all versus transverse coordinate
$x_2$. In all cases, the curved shape of the seed pulse is well
described by Eqns. (\ref{eq:selfsim2}) or (\ref{eq:ssbril2}).  Even
for strange pulse envelope shapes, like the ``donut'' shape of the
pulse with OAM, the relationships between pulse amplitude and duration
still hold. The presence or absence of parastic instabilities does not
have much of an influence on this behaviour. For example, in
Ref. \cite{trines11a}, one can see numerical simulations of two
Raman-amplified laser pulses, one with and one without filamentation.
Both pulses exhibit the horseshoe shape predicted here.

\end{document}